\documentclass[a4paper,11pt]{article}
\usepackage{pos}
\usepackage{hyperref}

\usepackage[utf8]{inputenc}
\usepackage{comment}

\usepackage{graphicx}%
\usepackage{dcolumn}%
\usepackage{bm}%
\usepackage{multibib}
\usepackage{xcolor}
\usepackage{wrapfig}

\newcommand{\ket}[1]{\ensuremath{| {#1} \rangle }}
\newcommand{\bra}[1]{\ensuremath{\langle {#1} |}}

\renewcommand{\vec}[1]{\bm{#1}}

\newcites{SM}{Supplementary Material Bibliography}

\title{
Lattice calculation of the $D_s\mapsto X \ell \bar{\nu}_\ell$ inclusive decay rate: an overview}

\author*[a,1]{A. De Santis}
\author*[b,1]{C. F. Gro\ss{}}

\affiliation[a]{Università di Roma Tor Vergata and INFN}

\affiliation[b]{HISKP and Bethe Centre for Theoretical Physics, University of Bonn, Germany}

\note{on behalf of the collaboration of
A. De Santis,
A. Evangelista,
R. Frezzotti,
F. Margari,
N. Tantalo from the University and INFN of Rome Tor Vergata,
P. Gambino,
M. Panero from the University of Turin,
G. Gagliardi,
V. Lubicz,
F. Sanfilippo,
S. Simula from the University and INFN of Roma Tre,
M. Garofalo,
C. F. Gro\ss{},
B. Kostrzewa,
C. Urbach from HISKP Uni Bonn and BCTP Uni Bonn,
A. Smecca from Swansea University
}

\emailAdd{alessandro.desantis@roma2.infn.it}
\emailAdd{gross@hiskp.uni-bonn.de}

\abstract{In this talks, on behalf of our collaboration
we present an overview of our first-principle lattice QCD calculation of the $D_s\mapsto X \ell \bar{\nu}_\ell$ inclusive decay rate. Here we introduce the theoretical background and focus on the methodological aspects of the calculation. A detailed discussion of our results, including the comparison with the corresponding experimental measurements will soon be presented in a forthcoming publication.\\

}

\FullConference{The 41st International Symposium on Lattice Field Theory (LATTICE2024)\\
 28 July - 3 August 2024\\
Liverpool, UK\\}

\renewcommand{\logo}{\relax} 

\begin{document}
\maketitle

\section{
\label{sec:introduction}
Introduction
}

In these talks we present, on behalf of our collaboration, an overview of our lattice QCD calculation of the $D_s\mapsto X \ell \bar{\nu}_\ell$ inclusive decay rate. From the phenomenological perspective, a first-principle lattice study of this process is important because it allows to use the experimental information of Refs.~\cite{CLEO:2009uah,BESIII:2021duu} to constrain the CKM matrix elements $V_{cs}$, $V_{cd}$ and $V_{us}$. From a theoretical perspective, our results are important because they show that inclusive semileptonic decays can nowadays be studied from first-principles on the lattice. The techniques we use for our calculation have been developed only recently~\cite{Hansen:2017mnd,Hashimoto:2017wqo,Hansen:2019idp,Gambino:2020crt,Gambino:2022dvu}. In a previous work~\cite{Gambino:2022dvu}, using the methods of~\cite{Hashimoto:2017wqo,Hansen:2019idp,Gambino:2020crt}, some of our collaborators have shown good agreement of the results obtained on the lattice for inclusive semileptonic decay rates with the OPE approach~\cite{Manohar:1993qn,Blok:1993va,Bigi:1992su,Bigi:1993fe,Chay:1990da}, that has been for many years the only viable theoretical approach to heavy meson inclusive semileptonic decays.
We have successfully used our methods in other
studies~\cite{ExtendedTwistedMassCollaborationETMC:2022sta,Evangelista:2023fmt,ExtendedTwistedMass:2024myu},
 while other groups have also started to tackle the problem of inclusive
semileptonic decays, see
Refs.\cite{Barone:2022gkn,Kellermann:2022mms,Barone:2023tbl,Kellermann:2023yec,Barone:2023iat,Kellermann:2024zfy,Hashimoto:2024pnd}.
 This presentation is focused on the methodological aspects of the calculation.
A detailed discussion of our results, including the comparison with the
corresponding experimental measurements will soon be presented in a forthcoming
publication.

\section{
\label{sec:contratemoments}
The differential decay rate
}
\begin{wrapfigure}{R}{0.33\textwidth}
\includegraphics[width=0.3\textwidth]{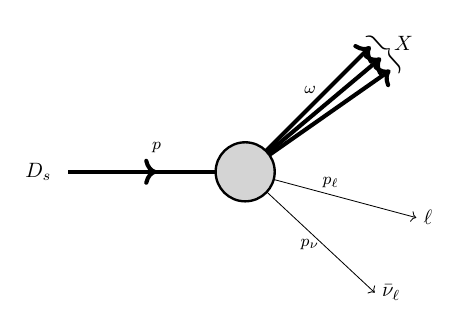}
\caption{\label{fig:kinematic} The kinematics of the inclusive $D_s\mapsto X\ell \bar \nu_\ell$ semileptonic decay. }
\end{wrapfigure}

In this work, we use a notation corresponding to~\cite{Gambino:2022dvu}. We work in the rest-frame of the decaying $D_s$ meson and call
\begin{flalign}
&
p=m_{D_s}(1,\vec 0)\;,
\qquad
\omega=m_{D_s}(\omega_0,\vec \omega)\;,
\qquad
p_\ell=m_{D_s}(e_\ell,\vec k_\ell)\;,
\qquad
p_\nu=m_{D_s}(e_\nu,\vec k_\nu)\;,
\end{flalign}
the four-momenta of the $D_s$, of the generic hadronic state $X$, of the lepton and of the neutrino, so that the energy-momentum conservation relation $p=p_\ell+p_\nu+\omega$ (see fig.~\ref{fig:kinematic}) implies
\begin{flalign}
\omega_0=1-e_\ell-e_\nu\;,
\qquad
\vec \omega=-\vec k_\ell -\vec k_\nu\;.
\end{flalign}
We work in the approximation in which the lepton is massless and therefore we have $\vec k_\ell^2=e_\ell^2$ as well as $\vec k_\nu^2=e_\nu^2$.

The fully inclusive process $D_s\mapsto X\ell \bar \nu_\ell$ can be separated into three different flavour channels, mediated by the flavour components $J^\mu_{\bar cs}$, $J^\mu_{\bar cd}$  and $J^\mu_{\bar us}$ of the hadronic weak current,
\begin{flalign}
&
J^\mu_{\bar fg}(x)
=\bar \psi_{\bar f}(x) \gamma^\mu(1-\gamma^5) \psi_g(x)\;.
\end{flalign}
We use the lower-case indexes $\bar f=\{\bar c,\bar c,\bar u\}$ and $g=\{s,d,s\}$ to indicate the flavour indexes of the currents.
When the flavour indexes are omitted, we refer to the fully inclusive process that is mediated by the sum of the three different flavour contributions weighted by the corresponding CKM matrix elements $V_{fg}$,
\begin{flalign}
J^\mu(x) =
V_{cs} J^\mu_{\bar cs}(x) +
V_{cd} J^\mu_{\bar cd}(x) +
V_{us} J^\mu_{\bar us}(x)\;.
\end{flalign}
Taking into account that the $D_s$ meson has $\bar s c$ flavour, in the channels mediated by $J^\mu_{\bar cd}, J^\mu_{\bar us}, J^\mu_{\bar cs}$ the final hadrons are $\bar s d, \bar u c$-flavoured and $\bar s s$-flavourless, respectively.

The currently available experimental results~\cite{CLEO:2009uah,BESIII:2021duu} provide the fully-inclusive decay rate $\Gamma\equiv \Gamma[D_s\mapsto X \ell \bar \nu_\ell]$ which is the sum,
\begin{flalign}
\Gamma=
\vert V_{cs}\vert^2 \Gamma_{\bar cs} +
\vert V_{cd}\vert^2 \Gamma_{\bar cd} +
\vert V_{us}\vert^2 \Gamma_{\bar us} \;,
\end{flalign}
of the contributions corresponding to the different flavour channels. In this work we provide separate results for the contributions of the dominant channel $\Gamma_{\bar cs}$ as well as for the Cabibbo-suppressed channel $\Gamma_{\bar cd}$. $\Gamma_{\bar us}$ is negligibly small.

Each contribution to the decay rate can be written as
\begin{flalign}
&
\Gamma_{\bar fg}=
G_F^2 S_\mathrm{EW} \int \frac{d^3 p_\nu}{(2\pi)^3}\,
\frac{d^3 p_\ell}{(2\pi)^3}\,
\frac{
L_{\mu\nu}(p_\ell,p_\nu) H_{\bar fg}^{\mu\nu}(p,p-p_\ell-p_\nu) }
{4 m_{D_s}^2 e_\ell e_\nu}\;,
\label{eq:gammfgstart}
\end{flalign}
where $G_F$ is the Fermi constant and $S_\mathrm{EW}=1.02$  accounts for the logarithmic electroweak correction \cite{Sirlin:1981ie}.
The leptonic tensor is given by a combination of the lepton and neutrino momentum
and the so-called hadronic tensor, which is in fact an hadronic spectral density, is given by
\begin{flalign}
H_{\mu\nu}(p,\omega)
=\frac{(2\pi)^4}{2m_{D_s}}
\bra{D_s}J_\mu^\dagger(0)\, \delta^4(\mathbb{P}-\omega)\, J_\nu(0)\ket{D_s}\;,
\end{flalign}
where $\mathbb{P}=(H,\vec P)$ is the isoQCD four-momentum operator. By relying
on Poincaré covariance, $H^{\mu\nu}(p,\omega)$ can be decomposed into five form
factors,
that, in our convention, are real and dimensionless. In the rest-frame of the $D_s$ meson the dependence of the form-factors upon the scalars $p\cdot \omega$ and $\omega^2$ can be traded for the dependence upon the variables $(\omega_0,\vec \omega^2)$.
In order to express the form-factors in terms of the different components of the hadronic tensor, we consider the two unit vectors $\vec{\hat n}_r$  orthogonal to $\vec{\hat \omega} = \vec \omega/\vert \vec \omega\vert$, i.e.
\begin{flalign}
\vec{\hat n}_r\cdot \vec{\hat n}_s=\delta_{rs}\;,
\qquad
\vec{\hat n}_r\cdot \vec{\hat \omega}=0\;,
\qquad
r,s=1,2\;,
\end{flalign}
and introduce the following quantities
\begin{flalign}
&
\mathcal{Y}^{(1)}=-\frac{m_{D_s}}{2}\sum_{r=1}^2\sum_{i,j=1}^3 \hat n^i_r \hat n^j_r H^{ij}(p,\omega)\;,
\nonumber
\qquad
\mathcal{Y}^{(2)}=m_{D_s}H^{00}(p,\omega)\;,
\nonumber\\[-5pt]
&
\mathcal{Y}^{(3)}=m_{D_s}\sum_{i,j=1}^3 \hat \omega^i \hat \omega^j H^{ij}(p,\omega)\;,
\nonumber
\qquad
\qquad
\mathcal{Y}^{(4)}=-m_{D_s}\sum_{i=1}^3 \hat \omega^i H^{0i}(p,\omega)\;,
\nonumber \\[-5pt]
&
\mathcal{Y}^{(5)}=-\frac{im_{D_s}}{2}\sum_{i,j,k=1}^3 \epsilon^{ijk} \hat \omega^k H^{ij}(p,\omega)\;,
\qquad
\mathcal{Y}^{(i)}\equiv \mathcal{Y}^{(i)}(\omega_0,\vec \omega^2)\;,
i=1,\cdots,5\;.
\label{eq:Ydefs}
\end{flalign}

The form factors can be expressed in terms of these quantities.
By relying on the form-factors decomposition, and by working out the phase-space kinematical constraints in the rest-frame of the $D_s$ meson, Eq.~(\ref{eq:gammfgstart}) can be rewritten as
\begin{flalign}
&
\Gamma_{\bar f g}=
G_F^2 S_\mathrm{EW}
\int_0^{(\vert\vec \omega\vert_{\bar F G}^\mathrm{max})^2} d\vec \omega^2
\int_{\omega_{\bar FG}^\mathrm{min}}^{\omega^\mathrm{max}} d\omega_0
\int_{e_\ell^\mathrm{min}}^{e_\ell^\mathrm{max}} de_\ell
\frac{d\Gamma_{\bar f g}}{d\vec \omega^2 d\omega_0 de_\ell}\;.
\label{eq:gammfgtripleint}
\end{flalign}
When the three integrations are performed in the order specified in Eq.~(\ref{eq:gammfgtripleint}), the integration limits to be used are given by the following expressions
\begin{flalign}
&
e_\ell^\mathrm{min}=\frac{1-\omega_0-\vert \vec \omega\vert}{2}\;,
\qquad
\omega_{\bar FG}^\mathrm{min}=\sqrt{r_{\bar FG}^2+\vec \omega^2}\;,
\qquad\vert\vec \omega\vert_{\bar FG}^\mathrm{max}= \left(\frac{1-r_{\bar FG}^2}{2}\right)\;,
\nonumber \\[-4pt]
&
e_\ell^\mathrm{max}=\frac{1-\omega_0+\vert \vec \omega\vert}{2}\;,
\qquad
\omega^\mathrm{max}=1-\sqrt{\vec \omega^2}\;,
\qquad\quad
r_{\bar FG} = \frac{m_{P_{\bar FG}}}{m_{D_s}}\, .
\label{eq:limits1}
\end{flalign}
\section{
\label{sec:contrate}
The total decay rate
}
In order to compute the total rate $\Gamma$, the integrals appearing in Eq.~(\ref{eq:gammfgtripleint}) have to be performed.
The form factors do not depend upon $e_\ell$, therefore
the lepton energy integral can be performed analytically, leaving
\begin{flalign}
&
\frac{1}{\bar \Gamma}
\frac{d \Gamma}{d \omega^0 d \vec \omega^2}
=
\vert \vec \omega\vert^3\, Z^{(0)}
+
\vert \vec \omega\vert^2 (\omega^\mathrm{max}-\omega_0)\, Z^{(1)}
+
\vert \vec \omega\vert (\omega^\mathrm{max}-\omega_0)^2\, Z^{(2)},
\label{eq:doubleGhs}
\end{flalign}
\begin{flalign}
\text{with}
\qquad
&
Z^{(0)}=\mathcal{Y}^{(2)}+\mathcal{Y}^{(3)}-2\mathcal{Y}^{(4)}\;,
\nonumber
\qquad
Z^{(1)}=2\left(\mathcal{Y}^{(3)}-2\mathcal{Y}^{(1)}-\mathcal{Y}^{(4)}\right)\;,
\nonumber \\[-5pt]
&
Z^{(2)}=\mathcal{Y}^{(3)}-2\mathcal{Y}^{(1)}\;,
\qquad
\qquad
\quad
\bar \Gamma= \frac{m_{D_s}^5 G_F^2 S_\mathrm{EW}}{48\pi^4}\;.
\label{eq:Zgamma}
\end{flalign}
The parity-breaking form factor $\mathcal{Y}^{(5)}/\vert \vec \omega \vert$ does not contribute to the total rate.

To compute the $\omega_0$ integral we first need to derive a mathematical representation of the decay rate that is suitable for a lattice evaluation. To this end, we start by introducing the kernels
\begin{flalign}
\Theta_\sigma^{(p)}(x) = x^p\, \Theta_\sigma(x)\;,
\label{eq:defthetap}
\end{flalign}
where $p$ is a variable that, in the following, will assume non-negative integer values while $\Theta_\sigma(x)$ is any Schwartz\footnote{i.e.\ infinitely differentiable and vanishing, together with all its derivatives, faster than any power of $x$ in the limit $x\mapsto \infty$} representation of the Heaviside function $\theta(x)$ which depends smoothly upon the smearing parameter $\sigma$ and which is such that
\begin{flalign}\label{eq:heaviside}
\lim_{\sigma\mapsto 0} \Theta_\sigma(x)=\theta(x)\;.
\end{flalign}
The introduction of this mathematical device allows to trade the $\omega_0$ phase-space integral,
with convolutions of the distributions $Z^{(p)}(\omega_0,\vec \omega^2)$ with smooth smearing kernels,
\begin{flalign}
&
\frac{1}{\bar \Gamma}
\frac{d \Gamma^{(p)}(\sigma)}{d \vec \omega^2}
=
\vert \vec \omega \vert^{3-p}\,
\int_{\omega^\mathrm{min}-\epsilon}^\infty d\omega_0\,
\Theta_\sigma^{(p)}(\omega^\mathrm{max}-\omega_0)\, Z^{(p)}(\omega_0,\vec \omega^2),
\label{eq:dGZint}
\end{flalign}
and with a limiting procedure,
\begin{flalign}
\Gamma
=
\sum_{p=0}^2
\int_0^{(\vert\vec \omega\vert^\mathrm{max}+\epsilon)^2} d\vec \omega^2\,
\lim_{\sigma\mapsto 0}
\frac{d\Gamma^{(p)}(\sigma)}{d \vec \omega^2}\;.
\label{eq:gammalimit}
\end{flalign}

We now rely on the Stone-Weierstrass theorem and observe that, for any positive value of the length scale $a$, the kernels $\Theta_\sigma^{(p)}(\omega^\mathrm{max}-\omega_0)$ can \emph{exactly} be represented according to
\begin{flalign}
\Theta_\sigma^{(p)}(\omega^\mathrm{max}-\omega_0)
=
\lim_{N\mapsto \infty}\sum_{n=1}^{N} g^{(p)}_n(N)\, e^{-\omega_0 (a m_{D_s}) n} \;.
\label{eq:stoneZ}
\end{flalign}
The coefficients $g^{(p)}_n(N)$ appearing in the previous formula have to be read as the coordinates of the kernels $\Theta_\sigma^{(p)}(\omega^\mathrm{max}-\omega_0)$ on the discrete set of basis-functions $\exp[-\omega_0 (a m_{D_s}) n]$.
On the lattice we cannot directly compute the distribution, but can compute the (amputated) Euclidean correlators
\begin{flalign}
\hat Z^{(p)}(t,\vec \omega^2)
=
\int_{\omega^\mathrm{min}-\epsilon}^\infty d\omega_0\, e^{-\omega_0 (m_{D_s}t)}\, Z^{(p)}(\omega_0,\vec \omega^2)
\label{eq:Zcorr}
\end{flalign}
at the discrete Euclidean times $t=an$, where $a$ is the lattice spacing. By using Eq.~(\ref{eq:stoneZ}) the connection can now easily be established,
\begin{flalign}\label{eq:lim_N_gamma}
\frac{1}{\bar \Gamma}
\frac{d \Gamma^{(p)}(\sigma)}{d \vec \omega^2}
=\vert \vec \omega \vert^{3-p}\,
\lim_{N\mapsto \infty}\sum_{n=1}^{N} g^{(p)}_n(N)\,
\hat Z^{(p)}(a n,\vec \omega^2)\; .
\end{flalign}

In order to determine the coefficients $g^p_n(N)$, and to study numerically the $N\mapsto \infty$ limit at fixed $\sigma>0$ and the associated systematic errors, we use the Hansen-Lupo-Tantalo (HLT) algorithm of Ref.~\cite{Hansen:2019idp}, see section~\ref{sec:hlt}.
Considerations of the $\sigma \mapsto 0$ limit are given in sec.~\ref{sec:sigmato0}, while the numerical integration of the $\vec \omega^2$ integral will be provided in the full paper.

\section{
\label{sec:sigmato0}
The \texorpdfstring{$\sigma\mapsto 0$}{s->0} limit
}
In the previous two sections, in order to compute the total rate and the leptonic moments on the lattice, we traded the compact $\omega_0$ phase-space integral with convolutions of the $Z^{(p)}(\omega_0,\vec \omega^2)$ distributions with the smooth kernels $\Theta_\sigma^{(p)}(\omega^\mathrm{max}-\omega_0)$ and with the $\sigma\mapsto 0$ limiting procedure. In order to understand how to perform numerically the required $\sigma\mapsto 0$ extrapolations we have to study the asymptotic behaviour for small values of $\sigma$
of
\begin{flalign}
d\Gamma^{(p)}(\sigma)/d \vec \omega^2
=
\int_{\omega^\mathrm{min}-\varepsilon}^\infty d\omega_0\,
\Theta^{(p)}_\sigma(\omega^\mathrm{max}-\omega_0)\, Z(\omega_0).
\label{eq:Oasympt1}
\end{flalign}
Careful analysis of this integral, especially its behaviour around the point $\omega^\mathrm{max}-\omega_0 \approx 0$ gives
\begin{flalign}
\frac{d\Gamma^{(0,1)}(\sigma)}{d \vec \omega^2}
&=
\frac{d\Gamma^{(0,1)}}{d \vec \omega^2}
+O(\sigma^2)\;,
\qquad
\frac{d\Gamma^{(2)}(\sigma)}{d \vec \omega^2}
=
\frac{d\Gamma^{(2)}}{d \vec \omega^2}
+O(\sigma^4)\;.
\label{eq:asymptG}
\end{flalign}
This explains our choice of organizing the calculation in terms of the kernels $\Theta^{(p)}_\sigma(\omega^\mathrm{max}-\omega_0)$ and, therefore, in terms of the distributions $Z^{(p)}(\omega_0,\vec \omega^2)$.
This leads to a faster rate of convergence for $p=2$.
\section{
\label{sec:correlators}
Lattice correlators
}
\begin{table*}[t]
\begin{center}
\begin{tabular}{lccccc}
ensemble & $L/a$ & $a~[\rm fm]$ & $L~[\rm fm]$ &  $Z_V$ & $Z_A$   \\
\hline
B48    & $48$    & $\quad$ $0.07957(13)$ $\quad$  & $\quad$ 3.82 $\quad$ & $\quad$ $0.706405(17)$  $\quad$  & $\quad$ $0.74267(17)$  $\quad$ \\
B64    & $64$    & $\quad$ $0.07957(13)$ $\quad$  & $\quad$ 5.09 $\quad$ & $\quad$ $0.706379(24)$  $\quad$  & $\quad$ $0.74294(24)$  $\quad$ \\
B96    & $96$    & $\quad$ $0.07957(13)$ $\quad$  & $\quad$ 7.64 $\quad$ & $\quad$ $0.706405(17)$  $\quad$  & $\quad$ $0.74267(17)$  $\quad$ \\[1pt]
C80    & $80$    & $\quad$ $0.06821(13)$ $\quad$  & $\quad$ 5.46 $\quad$ & $\quad$ $0.725404(19)$  $\quad$  & $\quad$ $0.75830(16)$  $\quad$ \\
[4pt]
D96    & $96$    & $\quad$ $0.05692(12)$ $\quad$ & $\quad$ 5.46 $\quad$ & $\quad$ $0.744108(12)$  $\quad$  & $\quad$ $0.77395(12)$  $\quad$ \\[1pt]
E112   & $112$   & $\quad$ $0.04891(6)$ $\quad$  & $\quad$ 5.48 $\quad$ & $\quad$ $0.758231(5)$  $\quad$  & $\quad$ $0.78542(7)$   $\quad$ \\
\hline
\end{tabular}
\caption{\normalfont{ETMC gauge ensembles used in this work. We give the values of the lattice spacing $a$, of the spatial lattice extent $L$, and of the vector and axial renormalization constants $Z_{V}$ and $Z_{A}$. The temporal extent of the lattice is always $T=2L$. }
\label{tab:iso_EDI_FLAG}}
\end{center}
\end{table*}
The lattice correlators needed to extract the decay rate have been computed on the physical-point gauge ensembles, listed in Table~\ref{tab:iso_EDI_FLAG}, that have been generated~\cite{Alexandrou:2018egz,ExtendedTwistedMass:2020tvp, ExtendedTwistedMass:2021qui,Finkenrath:2022eon} by the Extended Twisted Mass Collaboration (ETMC) with $n_f = 2 + 1 + 1$ flavours of Wilson-Clover Twisted Mass (TM) sea quarks. The bare parameters of the simulations have been tuned to match our scheme of choice for defining isoQCD, the so-called Edinburgh/FLAG consensus~\cite{FlavourLatticeAveragingGroupFLAG:2024oxs}.

We adopted the mixed-action lattice setup described in full details in the appendices of Ref.~\cite{ExtendedTwistedMassCollaborationETMC:2024xdf}. In this setup the action of the valence quarks is discretized in the so-called Osterwalder-Seiler (OS) regularization.
To interpolate the $D_s$ meson we use the following pseudoscalar operator
\begin{flalign}
P(x) = \bar \Psi_s(x) \gamma_5 \Psi_c(x) \;,
\end{flalign}
built in terms of the fields $\Psi_f$ that are obtained by applying Gaussian smearing to the fields $\psi_f$. The two-point correlator
\begin{flalign}\label{eq:twopoint}
C(t)=\sum_{\vec x} T\bra{0} P(t,\vec x)\, P^\dagger(0) \ket{0}
\end{flalign}
is used to amputate the four-points functions.
From $C(t)$ we extract the mass and the residue $R_P$ in the usual way.
The four-point correlators from which we extract the amputated correlators $\mathcal{\hat Y}^{(i)}(t,\vec \omega^2)$ are given by
\begin{flalign}
&
C_{\mu\nu}(t_\mathrm{snk},t,t_\mathrm{src},\vec \omega^2)
=
a^9\sum_{\vec x_\mathrm{snk},\vec x_\mathrm{src},\vec x}
e^{im_{D_s} \vec w\cdot \vec x}
T\bra{0}
P(x_\mathrm{snk}) J_\mu^\dagger(x) J_\nu(0) P^\dagger(x_\mathrm{src})\ket{0}\;,
\end{flalign}
where $x=(t,\vec x)$, $x_\mathrm{snk}=(t_\mathrm{snk},\vec x_\mathrm{snk})$ and  $x_\mathrm{src}=(t_\mathrm{src},\vec x_\mathrm{src})$, and the current is the lattice discretized version of the weak current. In our setup both the vector,
\begin{flalign}
&
V^\mu_{\bar fg}(x)
=Z_A \bar \psi_f(x) i\gamma^\mu\psi_g(x)\;,
\quad
\text{and axial,}
\quad
A^\mu_{\bar fg}(x)
=Z_V \bar \psi_f(x) i\gamma^\mu\gamma_5\psi_g(x)\;,
\end{flalign}
components of the weak current $J^\mu_{\bar fg}(x)=V^\mu_{\bar fg}(x)-A^\mu_{\bar fg}(x)$ are discretized in the so-called TM regularization. In the TM regularization  the vector current renormalizes with the ultraviolet-finite constant $Z_A$ while the axial current renormalizes with the finite constant $Z_V$. The values of $Z_V$ and $Z_A$ used in this calculation are given in Table~\ref{tab:iso_EDI_FLAG}.

\begin{wrapfigure}{R}{0.45\textwidth}
\centering
\includegraphics[width=0.43\textwidth]{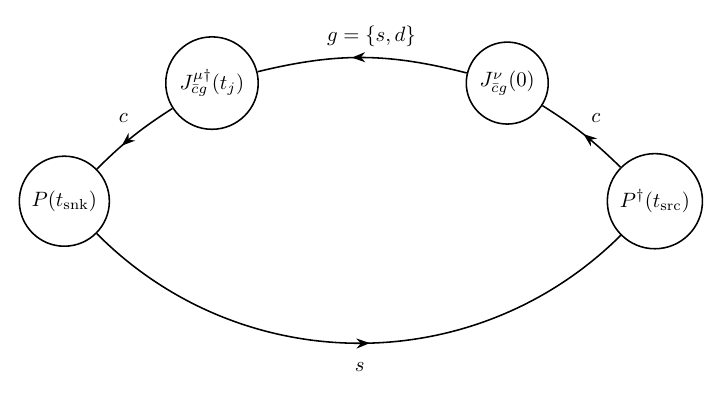}
\caption{\normalfont{Quark-connected Wick contraction contributing to $C^{\mu\nu}_{\bar c g}$ in the $\bar c s$ and $\bar c d$ channels. The spectator strange propagator arises from the contraction of the strange fields of the interpolating operators. The flavour-changing sequential propagator arises from the contraction of the charm fields of the interpolating operators with those of the currents and from the contractions of the $g=\{s,d\}$ down-type fields of the currents.}}
\label{fig:contdown}
\end{wrapfigure}
By integrating out the quark fields, the correlator $C^{\mu\nu}_{\bar f g}$ gets decomposed into the gauge-invariant contributions corresponding to the different fermionic Wick contractions.
The contractions corresponding to the quark-connected diagram shown in Figure~\ref{fig:contdown} contribute to both the dominating $\bar cs$ channel and to the Cabibbo suppressed $\bar cd$ channel and are obtained by contracting the charm quarks of the currents with those of the interpolating operators, the down-type quark of one current with that of the other current and the two strange quarks of the interpolating operators.

In the full paper we will show evidence
that the contribution from the remaining possible decay $\bar su$ channel, as well as a disconnected diagram contributing to the $\bar cs$ channel, are  small and negligible. We do not consider them in this work.

The asymptotic behaviour of the four-points correlator $C^{\mu\nu}(t_\mathrm{snk},t,t_\mathrm{src},\vec \omega^2)$ in the limits $T/2\gg t_\mathrm{snk} \gg t >0 \gg t_\mathrm{src} \gg -T/2$ is given by
\begin{flalign}
&
C^{\mu\nu}(t_\mathrm{snk},t,t_\mathrm{src},\vec \omega^2)
=
\frac{R_P}{4\pi m_{D_s}} e^{-m_{D_s}(t_\mathrm{snk}-t-t_\mathrm{src})}
\int_{\omega^\mathrm{min}-\epsilon}^\infty d\omega_0\, e^{-\omega_0 (m_{D_s}t)}\,
H^{\mu\nu}(p,\omega) +\cdots\;,
\label{eq:Cmunuasympt}
\end{flalign}
where $H_{\mu\nu}(p,\omega)$ is the hadronic tensor and the dots represent exponentially suppressed terms.
From the preceding definitions eq.~\ref{eq:twopoint} and eq.~\ref{eq:Ydefs}
we have extracted the correlators $\mathcal{\hat Y}^{(i)}(t,\vec \omega^2)$, e.g.
\begin{flalign}\label{eq:hat_Y}
&
\mathcal{\hat Y}^{(2)}(t,\vec \omega^2)
=
\lim_{t_\mathrm{snk}\mapsto \infty}\lim_{t_\mathrm{src}\mapsto -\infty}\lim_{T\mapsto \infty}
\frac{4\pi\, C^{00}(t_\mathrm{snk},t,t_\mathrm{src},\vec \omega^2)}
{R_P\, e^{-m_{D_s}(t_\mathrm{snk}-t-t_\mathrm{src})}}\;.
\end{flalign}
We determine the meson masses in the usual way and see good signals with not too much noise for all the ensembles from table~\ref{tab:iso_EDI_FLAG}.
From the analysis of $C(t)$ on the different ensembles we extracted the information needed to compute the correlators $C^{\mu\nu}(t_\mathrm{snk},t,t_\mathrm{src},\vec \omega^2)$ in the interesting region of the parameter space, i.e. for values of $t_\mathrm{src}$ and $t_\mathrm{snk}$ such that the systematic errors associated with the asymptotic limits $T\mapsto \infty$, $t_\mathrm{src}\mapsto -\infty$ and $t_\mathrm{snk}\mapsto \infty$ can be kept under control.

In order to extract the decay rate and the leptonic moments we used the data corresponding to the separation $t_\mathrm{snk}-t_\mathrm{src}\simeq 4.5$~fm, that we kept fixed in physical units on the different gauge ensembles. With this choice the systematics associated with the asymptotic limits can safely be neglected w.r.t. the statistical errors and, moreover, we can use larger values of $N$ to reconstruct the smearing kernels according to Eq.~(\ref{eq:stoneZ}) and, hence, to study the systematics associated with the $N\mapsto \infty$ limits (see section~\ref{sec:hlt}).

\section{
\label{sec:hlt}
The HLT algorithm and the \texorpdfstring{$N\mapsto \infty$}{N->inf} limit
}
In this section we provide an overview of the numerical implementation of the HLT algorithm~\cite{Hansen:2019idp} that we have used to extract the different contributions to the decay rate according to Eq.(\ref{eq:lim_N_gamma}).

We have considered two definitions of the smearing kernel $\Theta^{(p)}_\sigma(x)$ appearing in Eq.~(\ref{eq:defthetap}), based on
the following two regularizations of the Heaviside step-function,
\begin{flalign}
\Theta_\sigma(x)=\frac{1}{1+e^{-x/\sigma}} \;,
\quad \text{and} \quad
\Theta_\sigma(x)=\frac{1+\mathrm{erf}(x/\sigma)}{2}\; .
\label{eq:erf}
\end{flalign}
In the following we call ``sigmoid kernel'' and ``error-function kernel'' the smooth functions $\Theta^{(p)}_\sigma(x)$ obtained by multiplying for $x^{p}$ Eq.\ref{eq:erf}.
The two regularizations differ at $\sigma>0$ and become equivalent in the $\sigma\mapsto 0$ limit (see Eq.~\ref{eq:heaviside}). Moreover, by combining the numerical results corresponding to the two regularizations, we have a better control on the necessary $\sigma\mapsto 0$ extrapolations.


The coefficients $g_n^{(p)}(N)$ appearing in Eq.~(\ref{eq:stoneZ}) are obtained by minimizing the linear combination
\begin{flalign}
W_\alpha^{(p)}[\bm g]=\frac{A_\alpha^{(p)}[\bm g]}{A_\alpha^{(p)}[\bm 0]}+\lambda B^{(p)}[\bm g]
\label{eq:W_functional}
\end{flalign}
of the so-called norm functional
\begin{flalign}
&
A_\alpha^{(p)}[\vec g] = \int_{\omega^\mathrm{th}}^{\infty} d\omega_0\, e^{\alpha (am_{D_s})\omega_0} \
\left[ \Theta_\sigma^{(p)}(\omega^\mathrm{max}-\omega_0)
-\sum_{n=1}^{N}g_n e^{a(m_{D_s}\omega_0)n}\right]^2
\label{eq:norm_functional}
\end{flalign}
and of the error functional
\begin{flalign}\label{eq:error_functional}
B^{(p)}[\vec g]= \sum_{n_1,n_2=1}^{N} g_{n_1} g_{n_2}\mathrm{Cov}^{(p)}(an_1,an_2),
\end{flalign}
where the matrix $\mathrm{Cov}^{(p)}$ is the statistical covariance of the correlator $\hat Z^{(p)}(an,\bm{w}^2;a)$ at finite lattice spacing, and $N$ is bound from above by the number of time slices we have available for the four-point-correlator.
We set $\omega^\mathrm{th}=0.9\ \omega_{\bar F G}^\mathrm{min}$ and $\alpha=0$ in our analysis.
We measure the difference of the reconstruction to the kernel with
\begin{flalign}
d^{(p)}\left(N;\vec \Sigma\right)=\sqrt{\frac{A_0^{(p)}\left[\vec g^{(p)}(N;\vec \Sigma)\right]}{A_0^{(p)}[\vec 0]}}\;,
\end{flalign}

To study the dependence on the algorithmic parameters and obtain the optimal coefficients $\vec g^{(p)}_{\star}$ to evaluate Eq.~\eqref{eq:lim_N_gamma} we do the so-called HLT stability analysis already used in other
works~\cite{ExtendedTwistedMassCollaborationETMC:2022sta,Evangelista:2023fmt,ExtendedTwistedMass:2024myu,Frezzotti:2023nun}. The core of the HLT stability analysis is to search for independence with respect to the algorithmic parameters within statistical errors so that systematic effects can be safely estimated.
\begin{wrapfigure}{R}{0.45\textwidth}
\includegraphics[width=0.43\textwidth]{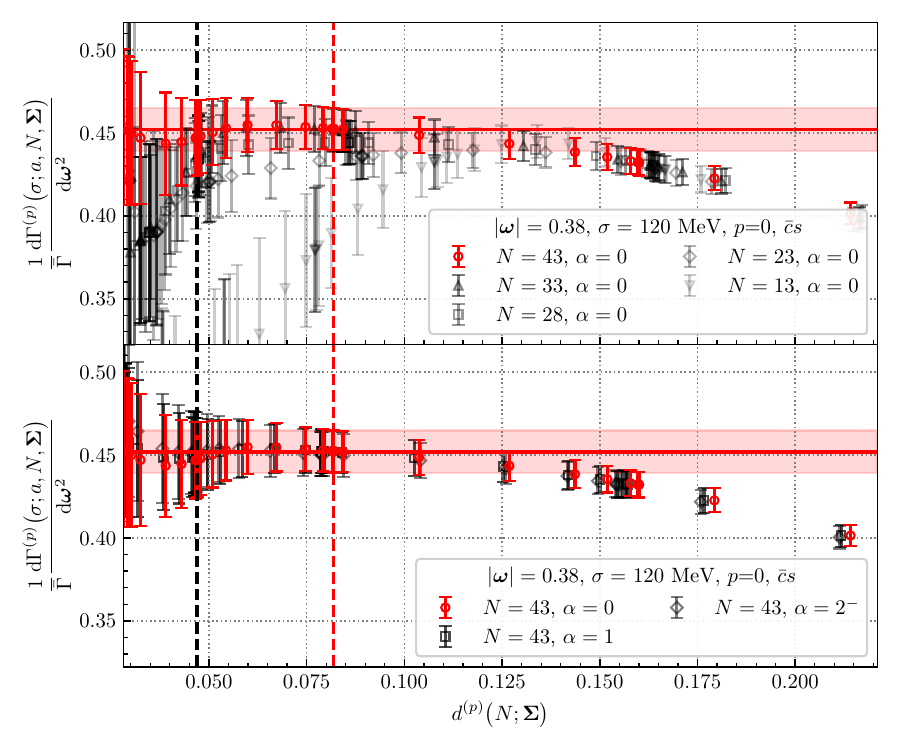}
\caption{
\normalfont{Stability analysis for the contribution $p=0$ to the total decay rate for the $\bar c s$  channel with smearing parameter
$\sigma=120$~MeV, spatial momentum $|\vec \omega|=0.38$, sigmoid kernel and D96 ensemble. Top panel: study of the limit $N\mapsto \infty$ by changing $\lambda$ at fixed $\alpha=0$. Bottom panel: study of the dependence on the definition of the norm functional  by changing $\lambda$ at fixed $N=43$.}
\vspace{-18mm}
}
\label{fig:stability_analysis}
\end{wrapfigure}
An example of a stability analysis is shown in Figure~\ref{fig:stability_analysis}.

In order to obtain a data-driven estimate of a systematic error on a given quantity $O$, for which we have different determinations $O_i$ that we expect to be equivalent in absence of a systematic error, we consider the pull variables and systematic errors
\begin{flalign}
\mathcal{P}^{ij}_\mathrm{sys} = \frac{O_i-O_j}{\Delta_{ij}}\;,
\qquad
    \Delta_\mathrm{sys}=
    \max_{ij}\left[
    \left|
    O_i-O_j
    \right|
    \mathrm{erf}\bigg(\frac{\mathcal{P}^{ij}_\mathrm{sys}}{\sqrt{2}}\bigg)
    \right]\; .
\label{eq:pull}
\end{flalign}
where $\Delta_{ij}$ is a conservative estimate of the error of the difference $O_i-O_j$ (depending upon the observable we consider either the error of one of the terms or the sum in quadrature of the errors of the two terms).
The error-function weights the difference with a (rough) estimate of the probability that the observed value is not due to a fluctuation.

In the case of the HLT stability analysis we estimate both the statistical errors and the central values of our results from the results at the  optimal $\vec g^{(p)}_{\star}$ point,
\begin{flalign}
\Delta^{(p)}_\mathrm{stat}(\vec \omega,\sigma)
\equiv
\Delta_\mathrm{stat}\left[
\frac{1}{\bar \Gamma}
\frac{d\Gamma^{(p)}_\star(\sigma)}{d\vec \omega^2}
\right]\;,
\end{flalign}
and the systematic error by calculating $\mathcal{P}^{(p)}_\mathrm{HLT}$ using
the results at $\vec g^{(p)}_{\star}$ and a corresponding $\vec
g^{(p)}_{\star\star}$ in Eq.~(\ref{eq:pull}).
In order to compute the decay rate and the lepton energy moments for each flavour channel, for all the considered values of $\vec \omega^2$ and of $\sigma$, on all the lattice ensembles and for the two different definitions of the smearing kernel (sigmoid and error-function), we performed more than 21000 stability analyses.
\section{
\label{sec:cs_DGammaDq2}
The $\Gamma_{\bar c s}$ and  $\Gamma_{\bar c d}$ contribution to the total decay rate
}

\begin{wrapfigure}{R}{0.45\textwidth}
\includegraphics[width=0.43\textwidth]{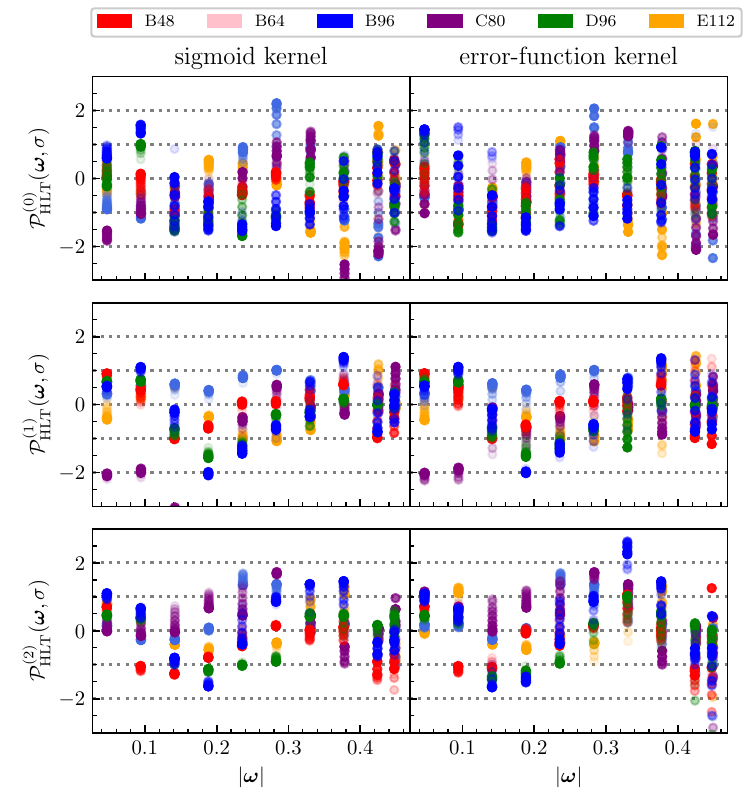}
\caption{\normalfont
Pull variable $\mathcal{P}_\mathrm{HLT}^{(p)}(\vec \omega,\sigma)$ for the
fermion-connected contributions  $d\Gamma^{(p)}_{\bar c s}(\sigma;a,L)/d\vec
\omega^2$. Different colours correspond to different ensembles while different
gradations of the same colour correspond to different values of $\sigma$.}
\vspace{-5mm}
\label{fig:cs_pull_HLT}
\end{wrapfigure}
In this section, we discuss our results for the fermion-connected contribution to the dominant $\Gamma_{\bar c s}$ channel, as well as the $\Gamma_{\bar c d}$ channel.

When the fermion-connected contribution is separated from the weak-annihilation contribution
the lightest possible hadronic state $P_{\bar s s}$ is not the neutral pion but the unphysical $\eta_{\bar s s}$ meson
, because no single neutral pion can be generated from the sea.
We have extracted the mass of the  $\eta_{\bar s s}$ meson from the quark-connected contribution to the appropriate two-point function as defined in eq.~\ref{eq:twopoint}.
We determined $r_{\bar s s-\mathrm{conn}}$ and
have then computed the fermion-connected Wick contraction of the correlators $C^{\mu\nu}_{\bar c s}(t_\mathrm{snk},t,t_\mathrm{src},\vec \omega^2)$ for  10 values of $\vert\vec \omega\vert$ ranging from 0.05 to 0.45. The values of the smearing parameters we use in the $\sigma \mapsto 0$ extrapolation are in the range $[10,200]$~MeV for the sigmoid kernel and $[100,700]$~MeV for the error-function kernel.

We show in Figure~\ref{fig:cs_pull_HLT} the pull variable $\mathcal{P}^{(p)}_\mathrm{HLT}(\vec \omega,\sigma)$.
It
shows we have $\mathcal{P}^{(p)}_\mathrm{HLT}(\vec \omega,\sigma)<3$ and only in a very few cases $\mathcal{P}^{(p)}_\mathrm{HLT}(\vec \omega,\sigma)>2$. This means that all our results are in the statistically dominated regime, i.e.\ the HLT systematic error $\Delta_\mathrm{HLT}^{(p)}(\vec \omega,\sigma)$ is smaller than the statistical error $\Delta_\mathrm{stat}^{(p)}(\vec \omega,\sigma)$.

\begin{wrapfigure}{R}{0.45\textwidth}
\includegraphics[width=0.43\textwidth]{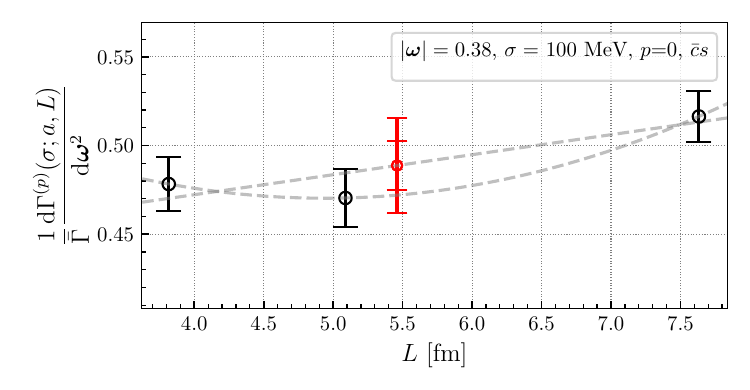}
\caption{\normalfont{
Interpolation of the results $d\Gamma^{(0)}_{\bar c s}(\sigma;a_B,L)/d\vec \omega^2$ at the reference volume $L_\star\simeq 5.5$~fm. The red point is the result of the linear interpolation and the larger error bar takes into account our estimate of the FSE systematic error.}}
\label{fig:cs_FSE}
\end{wrapfigure}

In order to estimate the finite size effects systematic errors $\Delta_L^{(p)}(\vec \omega,\sigma)$ we used the three ensembles B48, B64 and B96 at the coarsest simulated value of the lattice spacing (see Table~\ref{tab:iso_EDI_FLAG}). While the ensembles C80, D96 and E112
have been generated at the same reference physical volume,
the volumes of the B48, B64 and B96 ensembles are different.
In Figure~\ref{fig:cs_FSE} we illustrate our procedure to interpolate to the common volume and to estimate the FSE systematic errors. We perform linear and quadratic interpolations of the results of the three B-ensembles.
From these fits we obtain $d\Gamma^{(p)}(\sigma;a_B,L_\star)/d\vec \omega^2$ by taking the central value from the linear fit and by adding in quadrature to the error of the linear interpolation a systematic error estimated from the spread between the linear and the quadratic interpolation according to Eq.~(\ref{eq:pull}).

We then estimate the FSE systematic errors on our results $d\Gamma^{(p)}\big(\sigma\big)/d\vec \omega^2$, introducing
\begin{flalign}\label{eq:PFSE}
\mathcal{P}^{(p)}_\mathrm{FSE}(\vec \omega,\sigma)=
\frac{
\frac{d\Gamma^{(p)}\big(\sigma;a_B,L_\star\big)}{d\vec \omega^2}
    -
    \frac{d\Gamma^{(p)}\big(\sigma;a_B,7.6~\mathrm{fm}\big)}{d\vec \omega^2}
}{\Delta^{(p)}_\mathrm{stat}(\vec \omega,\sigma;a_B,L_\star)\, \bar \Gamma}
\,,
\end{flalign}
where $d\Gamma^{(p)}\big(\sigma;a_B,7.6~\mathrm{fm}\big)/d\vec \omega^2$ is the B96 result.
We use the same estimate of $\Delta_L^{(p)}(\vec \omega,\sigma)$ for all simulated values of the lattice spacing. We will show the values of the pull variable $\mathcal{P}^{(p)}_\mathrm{FSE}(\vec \omega,\sigma)$ in the full paper.
We see that
in all cases we have $\mathcal{P}^{(p)}_\mathrm{FSE}(\vec \omega,\sigma)<2$ and in most of the cases $\mathcal{P}^{(p)}_\mathrm{FSE}(\vec \omega,\sigma)<1$. This means that the FSE systematic errors on our results are much smaller than the corresponding statistical errors.
\begin{wrapfigure}[20]{R}{0.45\textwidth}
\centering
\includegraphics[width=0.43\textwidth]{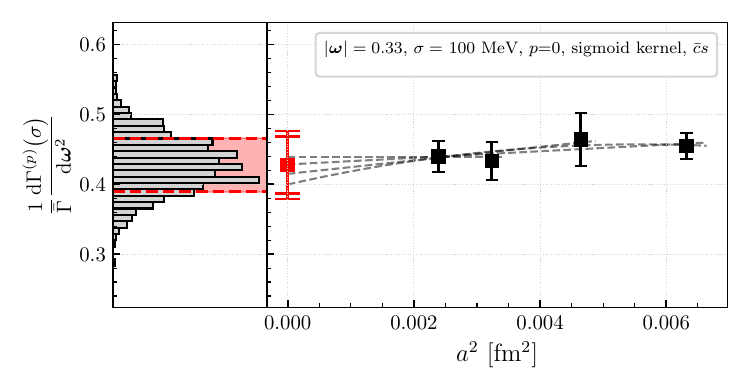}
\caption{\normalfont{Continuum extrapolation of the fermion-connected $d\Gamma^{(0)}_{\bar c s}(\sigma;L_\star)/d\vec \omega^2$ to the decay rate. The data correspond to $|\vec \omega|=0.44$, to $\sigma=100$~MeV and to the sigmoid smearing kernel. The different four dashed lines correspond to the different fits that we combine by using the Bayesian model average procedure discussed in the text. The histogram shows the distribution of the weighted bootstrap samples, the horizontal red dashed lines are the 16\% and 84\% percentiles while the red band is the statistical error.  The red point is the continuum result with the larger error bar taking into account our estimate of the systematic error associated with FSE. }}
\label{fig:cs_a}
\end{wrapfigure}

In Figure~\ref{fig:cs_a} we show an example of the continuum extrapolation of our results $d\Gamma^{(p)}\big(\sigma;a,L_\star\big)/d\vec \omega^2$.  We have four points (see Table~\ref{tab:iso_EDI_FLAG}) and we perform four different extrapolations: a constant fit of the two finer points (ensembles E112; D96); a fit linear in $a^2$ of the three finer points (ensembles E112, D96, C80); a fit linear in $a^2$ and a fit quadratic in $a^2$ of all the points. The different fits are combined by employing the Bayesian model average~\cite{Akaike:1974vps} that we have already employed in Ref.~\cite{Evangelista:2023fmt}.

Given $N$ different fits, the central value of the extrapolated result is given by
\begin{flalign}
    \bar{x} = \sum_{k=1}^{N} \omega_k x_k,
   \omega_k \propto
   \exp\big[-(\chi^2_k+2N^k_\mathrm{params}-N^k_\mathrm{points})/2\big]\;,
    \label{eq:AICaverage2}
\end{flalign}
where $x_k$ are the extrapolated results of each separate fit, the weights
$\omega_k$ are normalised, $\chi^2_k$, $N^k_\mathrm{params}$,
$N^k_\mathrm{points}$ are the $\chi^2$-variable, the number of parameters and
the number of points of the different fits. The total error is estimated by
using
\begin{flalign}
    \Delta^2_\mathrm{tot} = \sum_{k=1}^{N}\omega_k\Delta_k^2 +  \sum_{k=1}^{N}\omega_k(x_k-\bar x)^2\;,
    \label{eq:AICaverage3}
\end{flalign}
where the first sum is the weighted average of the square of the errors $\Delta_k$ on $x_k$ coming from the different fits while the second sum
provides an estimate for the systematic error. We employ the same procedure to extrapolate our results to the $\sigma \mapsto 0$ limit (see below).

As an additional quality indicator we use the pull-variable
$
\displaystyle
\mathcal{P}_a = \frac{\left\vert \bar x -x(a_E)\right\vert}{\Delta_\mathrm{tot}},
$
where $x(a_E)$ is the result at the finest value of the lattice spacing (here the E112 ensemble).
In more than $95\%$ of the cases, $\mathcal{P}_a$ is less than one, and we never observe $\mathcal{P}_a > 2$.
In more than $90\%$ of the cases, $\chi^2/\mathrm{d.o.f}\le 1$ for the dominant fit, and in more than $80\%$ of the cases, the dominant fit has only one parameter.
This is quantitative proof that our continuum limit is very flat and stable, and makes us very confident on the quality of our continuum extrapolations.

After performing the continuum extrapolations we add in quadrature to the error of the continuum results (including our estimates of the HLT and continuum-extrapolation systematic uncertainties) our estimates of the FSE systematic errors.
We can neglect the volume dependence and write $d\Gamma^{(p)}(\sigma)/d\vec \omega^2$.

\begin{wrapfigure}{R}{0.45\textwidth}
\includegraphics[width=0.43\textwidth]{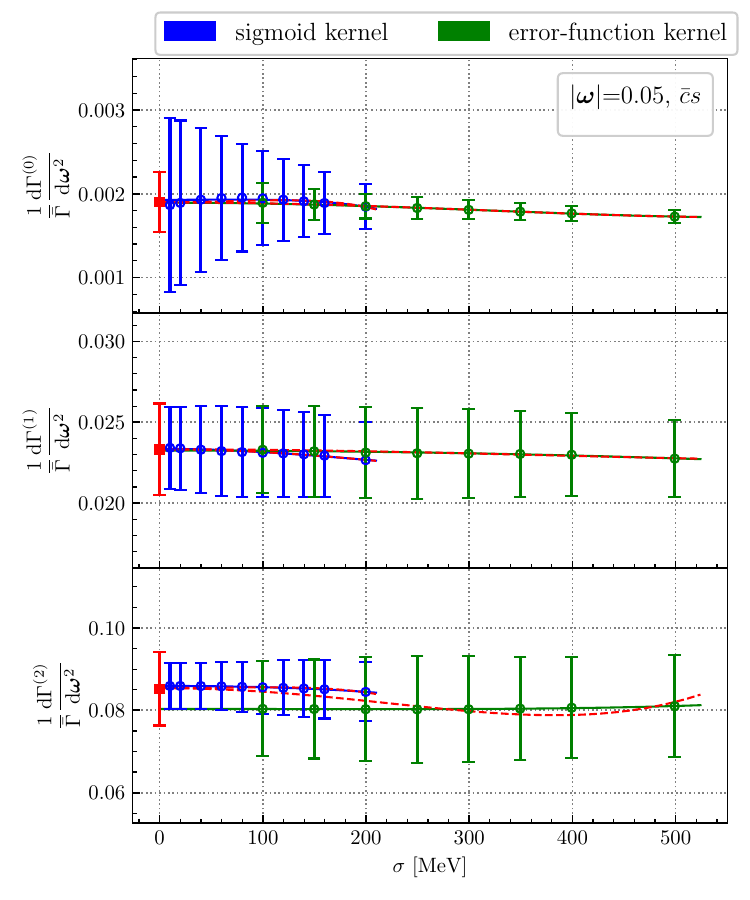}
\vspace{-3mm}
\caption{\normalfont{$\sigma \mapsto 0$ extrapolation of the connected $d\Gamma^{(p)}_{\bar c s}/d\vec \omega^2$ contribution to the differential decay rate. The data correspond to $\vert \vec \omega\vert=0.05$. The blue and green solid lines are the separate fits of the results obtained by using respectively the sigmoid and the error-function smearing kernels. The red line is the combined fit of both data sets. The red point is the final extrapolated result.} }
\vspace{-14mm}
\label{fig:cs_sigma_1}
\end{wrapfigure}

After having performed the continuum extrapolations, we perform the necessary $\sigma\mapsto 0$ extrapolations. To this end, we use the asymptotic formulae  of Eq.~(\ref{eq:asymptG}) and, for each fermion-connected contribution $d\Gamma^{(p)}_{\bar c s}(\sigma)/d\vec \omega^2$, we perform a polynomial fit. All fits contain only even powers of $\sigma$, for $p=0, 1$, the powers are $0, 2, 4$, for $p=2$ the powers are $0, 4, 6$.
We combine fits of the sigmoid data, the erf data, and a constrained fit with Eq.~\ref{eq:AICaverage2},~\ref{eq:AICaverage3}
to obtain our estimates of the connected contributions $d\Gamma^{(p)}_{\bar c s}/d\vec \omega^2$ to the physical differential decay rate.
Examples of these extrapolations are shown in Figure~\ref{fig:cs_sigma_1},
corresponding to a point close to the lower-end of the phase-space integration interval $[0,\vert\vec \omega\vert_{\bar ss-\mathrm{conn}}^\mathrm{max}=0.44]$. %
In this kinematic configuration our results show a very mild dependence upon $\sigma$, almost negligible within the errors that, at this stage, include our estimates of the systematics associated with the HLT stability analysis, with FSE and with the continuum extrapolations.
In the case of larger smearing parameters, the extrapolation is less flat, but consistent with the expected behaviour.
At the smaller considered values of $\sigma$ there is no significant difference between the results of the two smearing kernels.
We also look at an additional pull-variable for the robustness of this extrapolation, and it is almost always $<0.5$.
These facts make us confident on the robustness of our extrapolations.
\begin{figure}[hbt]
\begin{center}
    \includegraphics[width=0.4\linewidth]{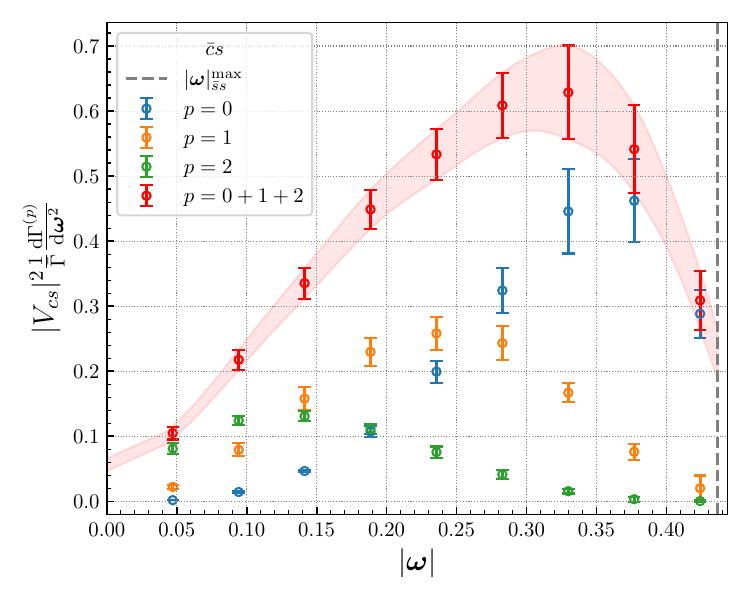}
    \includegraphics[width=0.4\linewidth]{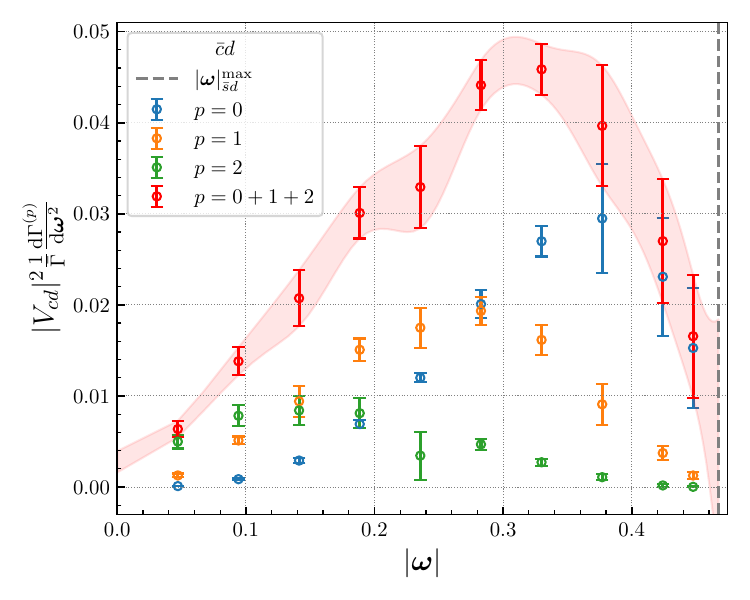}
\end{center}
    \caption{\normalfont{Left: Fermion-connected contribution $d\Gamma_{\bar c
    s}/d\vec \omega^2$ to the physical differential decay. The red points
    correspond to the sum of the three contributions $d\Gamma^{(p)}_{\bar c
    s}/d\vec \omega^2$ that are also shown in different colours. The error-bars
    correspond to the total error, i.e.\ to the sum in quadrature of the
    statistical errors and of the HLT, FSE, $a\mapsto 0$ and $\sigma\mapsto 0$
    systematic errors.
    Right: analogous for $d\Gamma_{\bar c d}/d\vec \omega^2$}}
    \label{fig:cs_final}
\end{figure}
Our final results for the fermion-connected contribution $d\Gamma_{\bar c s}/d\vec \omega^2$ to the physical differential decay rate are shown on the left-hand side of Figure~\ref{fig:cs_final}, the right-hand side shows the results for $d\Gamma_{\bar c d}/d\vec \omega^2$ obtained with the same analysis procedure.

\section{
\label{sec:outlook}
Outlook
}

In order to compare the results of section~\ref{sec:cs_DGammaDq2} to the experimental results, we interpolate the results with splines and perform the integral numerically.
An analogous procedure can be done for the first and second lepton energy moment, all of which will be reported in the forthcoming paper.

\vspace {-2mm}
\acknowledgments
The authors thank the other collaboration members:
A. Evangelista,
R. Frezzotti,
F. Margari,
N. Tantalo from the University and INFN of Rome Tor Vergata,
P. Gambino,
M. Panero from the University of Turin,
G. Gagliardi,
V. Lubicz,
F. Sanfilippo,
S. Simula from the University and INFN of Roma Tre,
M. Garofalo,
B. Kostrzewa,
C. Urbach from HISKP Uni Bonn and BCTP Uni Bonn,
A. Smecca from Swansea University.

The authors gratefully acknowledge the Gauss Centre for Supercomputing e.V. (www.gauss-centre.eu) for funding this project by providing computing time on the GCS Supercomputer JUWELS~\cite{JUWELS} at Jülich Supercomputing Centre (JSC), and the granted access to the Marvin cluster hosted by the University of Bonn.
This work has been supported as part of NRW-FAIR by the MKW NRW under the funding code NW21-024-A and by the Italian Ministry of University and Research (MUR) under the grant PNRR-M4C2-I1.1-PRIN 2022-PE2 Non-perturbative aspects of fundamental interactions, in the Standard Model and beyond F53D23001480006 funded by E.U.-NextGenerationEU.

\vspace{50mm}

\newpage
\bibliographystyle{custom_lattice_2024_gross}
\bibliography{incds}%

\end{document}